\documentclass[aps,prl,twocolumn,superscriptaddress]{revtex4-1}
\usepackage[colorlinks=true,linkcolor=blue,anchorcolor=red,citecolor=blue,urlcolor=blue]{hyperref}
\usepackage{bm}
\usepackage{graphicx}
\usepackage{color}
\usepackage{amsmath}
\usepackage{amssymb}

\makeatletter

\def\Eq#1{Eq.~(\ref{#1})}

\def\Fig#1{\text{Fig.}~\ref{#1}}

\begin{document}
	
	\title{Observation of Giant Nernst plateau in ideal 1D Weyl Phase}

\author{Y. Zhang\footnotemark[a]}
\thanks{These authors have contributed equally to this work.}
\affiliation{Anhui Key Laboratory of Condensed Matter Physics at Extreme Conditions, High Magnetic Field Laboratory, HFIPS, Chinese Academy of Sciences, Hefei 230031, China}

\author{Q. Li}
\thanks{These authors have contributed equally to this work.}
\affiliation{Anhui Key Laboratory of Condensed Matter Physics at Extreme Conditions, High Magnetic Field Laboratory, HFIPS, Chinese Academy of Sciences, Hefei 230031, China}

\author{Peng-Lu Zhao}
\thanks{These authors have contributed equally to this work.}
\email{zhaoplu@gmail.com}
\affiliation{Department of Physics, University of Science and Technology of China, Hefei, Anhui 230026, China}
\affiliation{Department of Physics, Southern University of Science and Technology (SUSTech), Shenzhen 518055, China}

\author{Y. C. Qian}
\affiliation{Anhui Key Laboratory of Condensed Matter Physics at Extreme Conditions, High Magnetic Field Laboratory, HFIPS, Chinese Academy of Sciences, Hefei 230031, China}

\author{Y. Y. Lv}
\affiliation{National Laboratory of Solid State Microstructures, Nanjing University, Nanjing 210093, China}

\author{Y. B. Chen}
\affiliation{National Laboratory of Solid State Microstructures, Nanjing University, Nanjing 210093, China}

\author{Q. Niu}
\affiliation{Department of Physics, University of Science and Technology of China, Hefei, Anhui 230026, China}
\affiliation{CAS Key Laboratory of Strongly-Coupled Quantum Matter Physics,
University of Science and Technology of China, Hefei, Anhui 230026, China}

\author{Hai-Zhou Lu}
\email{luhz@sustech.edu.cn}
\affiliation{Department of Physics,
Southern University of Science and Technology (SUSTech), Shenzhen 518055, China}
\affiliation{Quantum Science Center of Guangdong-Hong Kong-Macao Greater Bay Area (Guangdong), Shenzhen 518045, China}
\affiliation{International Quantum Academy, Shenzhen 518048, China}

\author{J. L. Zhang}
\email{zhangijnglei@hmfl.ac.cn}
\affiliation{Anhui Key Laboratory of Condensed Matter Physics at Extreme Conditions, High Magnetic Field Laboratory, HFIPS, Chinese Academy of Sciences, Hefei 230031, China}

\author{M. L. Tian}
\email{tianml@hmfl.ac.cn}
\affiliation{Anhui Key Laboratory of Condensed Matter Physics at Extreme Conditions, High Magnetic Field Laboratory, HFIPS, Chinese Academy of Sciences, Hefei 230031, China}
\affiliation{School of Physics and Materials Sciences, Anhui University, Hefei 230601, Anhui,China}

	
\begin{abstract}

The search for a giant Nernst effect beyond conventional mechanisms offers advantages for developing advanced thermoelectric devices and understanding charge-entropy conversion. Here, we study the Seebeck and Nernst effects of HfTe$_{5}$ over a wide range of magnetic fields. By tracking the unusual magneto-thermoelectric responses, we reveal two magnetic-field-driven phase transitions proposed for weak topological insulators: the gap-closing transition of the zeroth Landau bands and the topological Lifshitz transition. After the magnetic fields exceed approximately ten times the quantum limit, we observe that the Nernst signal no longer varies with the fields, forming a plateau with a remarkably large value, reaching up to $50~\mu\mathrm{V}/\mathrm{K}$ at $2~\mathrm{K}$. We theoretically explain the giant Nernst plateau as a unique signature of the ideal 1D Weyl phase formed in such high fields. Our findings expand the understanding of ideal Weyl physics and open new avenues for realizing novel thermoelectric effects without fundamental constraints.

\end{abstract}
	
	\maketitle


\noindent 
The Nernst effect is the transverse electric field $\textbf{E}$ produced by a longitudinal thermal gradient $\nabla T$ in the presence of a magnetic field. In recent years, it has been extensively investigated because of its physical mechanism and thermoelectric applications \cite{Behnia16RPP}. The orthogonal relationship between $\textbf{E}$ and $\nabla T$ enables a more efficient conversion of heat into electrical energy than that of the conventional longitudinal geometry \cite{Bell08S,Sakuraba13APE}. One of the representative transverse thermoelectric effects is the anomalous Nernst effect (ANE), where $S_{xy}$ initially increases, saturates, and forms a plateau in a range of weak magnetic fields. Recently, it has been recognized that the large Berry curvature, originating from Bloch electronic bands with spin-orbit coupling, can generate a significant transverse response in magnetic Weyl fermions \cite{Nagaosa10RMP,Nakatsuji15S,Muhammad17NP}. For instance, the ANE of Co-based Heusler alloys is about $6~\mu\mathrm{V}/\mathrm{K}$ at room temperature \cite{Sakai18NP}, and UCo$_{0.8}$Ru$_{0.2}$Al has recently been reported to exhibit a colossal ANE of $23~\mu\mathrm{V}/\mathrm{K}$ \cite{Asaba21SA}. However, the transverse thermoelectric effect based on ANE is several orders of magnitude smaller than the longitudinal thermoelectric effect. Moreover, it decays drastically with decreasing temperature. Given that the ANE occurs in weak magnetic fields with electrons following magnetic Bloch bands, one may ask whether strong magnetic fields, where electrons follow Landau bands, can offer new opportunities to overcome these limitations and achieve a larger Nernst effect.

\begin{figure*}[tb]\centering
		\includegraphics[width=17cm]{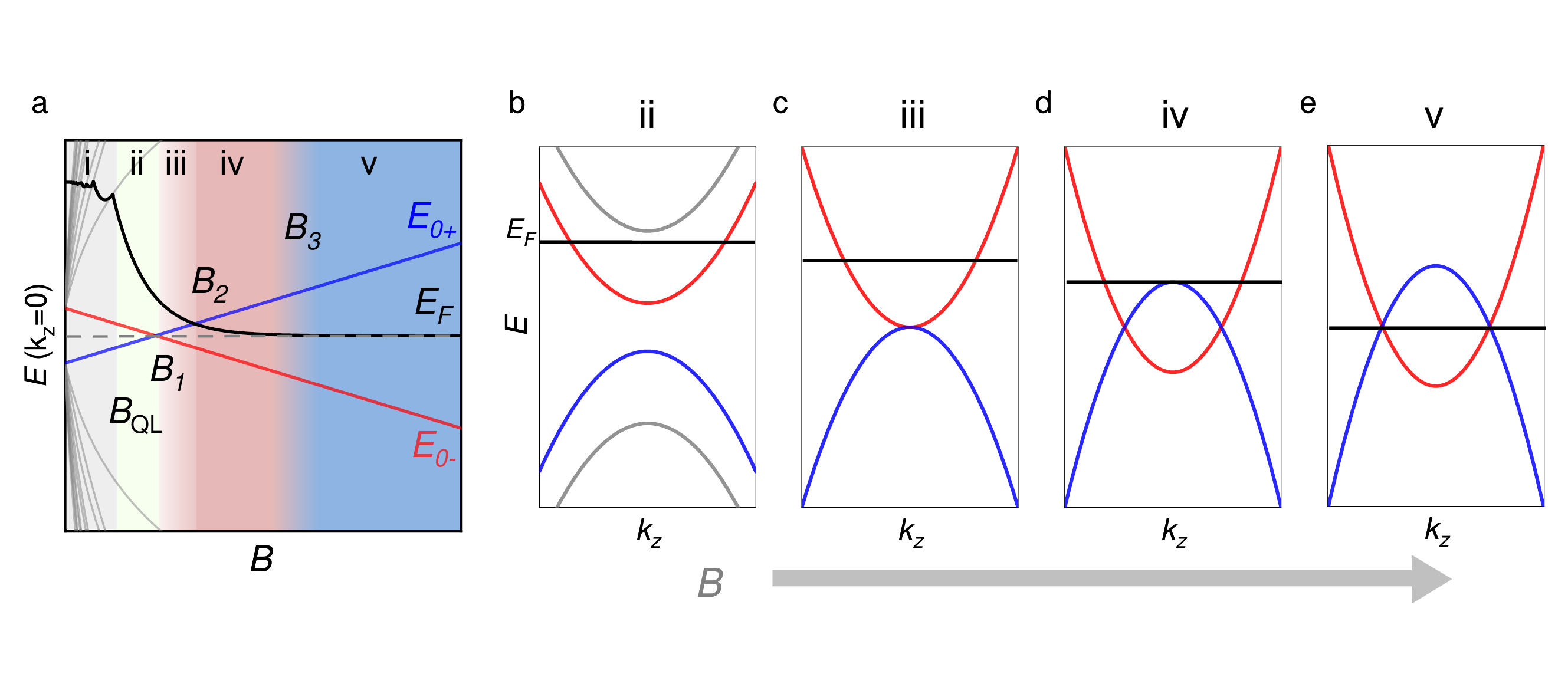}
		\caption{\textbf{Sketch of magnetic-field-induced Landau band transitions.} (a), Landau bands at $k_z=0$ and Fermi level ($E_{F}$) for a 3D weak topological insulator versus the magnetic field. The red, blue, and gray lines represent the zeroth Landau bands $E_{0+}$ (electron), $E_{0-}$ (hole), and high-index Landau bands, respectively. The black line represents the Fermi energy $E_{F}$. The i-v regions represent different phases: i) the 3D weak topological insulator state, ii) the system reaches the quantum limit, iii) the 1D trivial semiconductor state, iv) the 1D Weyl phase, and v) the ideal 1D Weyl phase. (b)-(e), Evolution of the Landau bands for weak topological insulators under selected magnetic fields. (b), Fermi level crosses only the zeroth Landau band at the quantum limit $B_{\mathrm{QL}}$. (c), Gap of the zeroth Landau band closes at the critical field $B_{1}$. (d), Topological Lifshitz transition occurs at $B_{2}$, where the Fermi level crosses the bottom of $E_{0-}$. (e), Ideal 1D Weyl states emerge after $B_{3}$, with all Weyl points located at the Fermi energy, $E_{F}$.}\label{fig.1}
	\end{figure*}

Various exotic quantum phenomena under strong magnetic fields have been discovered \cite{Lil08Sci,Fauque13L,Zhuzw17NC,Wangjh20PNAS} when the system reaches its quantum limit ($B_\mathrm{QL}$), where only the lowest Landau band is occupied. In the past decade, 3D topological materials have also opened up a new territory along this direction \cite{Liangt13NC,Moll16NC,LiuYW,Zhangcl17NP,Chenzg17PNAS,Assaf17L,Ramshaw18NC,Wanghc18SA,Zhangcl19NC,Tangfd19N,Liangsh19NM,Zhangjl19L,Zhangwj20NC,Zhangcl23arXiv}. 
In particular, for 3D weak topological insulators, two intriguing transitions within the lowest Landau bands have been proposed \cite{Wuwb23NM,Galeski22NC}. As illustrated in \Fig{fig.1}a, the two gapped lowest Landau bands become gapless at the critical magnetic field $B_{1}$ due to band inversion and Zeeman effects. This gap-closing transition can be understood analytically, as the lowest Landau bands are generally described by the following expression (see Supplementary Section I):
	\begin{equation}
		E_{0\xi}=-\xi \left( M_{z}k_{z}^{2}-\alpha B+\Delta \right),\label{EqLLBxi}
	\end{equation}
where $\xi = \pm 1$ represents the spin index, $M_{z}$ and $M_{\perp}$ are parameters related to band inversion, $\alpha = g\mu_{B}/2 - M_{\perp}e/\hslash$, with $g$ as the $g$-factor, and $\Delta$ is the bulk gap. For a weak topological insulator, $\Delta > 0$, $M_{\perp} < 0$, and $M_{z} > 0$, leading to $B_1 = \Delta/\alpha$. By further increasing $B$ to $B_{2}$, $E_{F}$ intersects the top of the energy band $E_{0-}$ (see \Fig{fig.1}d), triggering a topological Lifshitz transition where the Fermi surface shifts from two 1D Weyl points with the same spin to three with different spins. So far, experimental evidence for these transitions remains scarce \cite{Wuwb23NM,Galeski22NC}, and no evidence from thermoelectric transport has been found.
More importantly, further increasing $B$ beyond $B_3$ will push the Fermi energy into the 1D Weyl points (see \Fig{fig.1}f), forming an 'ideal 1D Weyl phase', for which experimental exploration is entirely absent.


In this work, we report the observation of a giant Nernst plateau in the ultra-quantum limit ($B>10B_\mathrm{QL}$) of hafnium pentatelluride HfTe$_{5}$, a weak topological insulator \cite{liu2024nc,Wuwb23NM}, reaching $50~\mu\mathrm{V}/\mathrm{K}$ at $2~\mathrm{K}$ (see \Fig{fig.3}b). Such a plateau has never been reported before and is unexpected within the conventional picture of Berry curvature. Further theoretical analysis reveals that the Nernst plateau originates from the formation of the ideal 1D Weyl phase. Prior to reaching the plateau, we observe distinct thermoelectric signatures corresponding to the gap-closing transition at $B_{1}$ and the topological Lifshitz transition at $B_{2}$. The magneto-thermopower converging to zero, along with the sign change in the Nernst signal, clearly indicates the gap-closing transition. The dome-like behavior of $-S_{xx}$ in the quantum limit signals the onset of the topological Lifshitz transition, with its maximum precisely determining $B_{2}$. Our high-field thermoelectric studies provide a more definitive means to elucidate these transitions, as suggested by previous experiments \cite{Wuwb23NM,Galeski22NC}.

	
\begin{figure*}[tb]\centering
		\includegraphics[width=17.5cm]{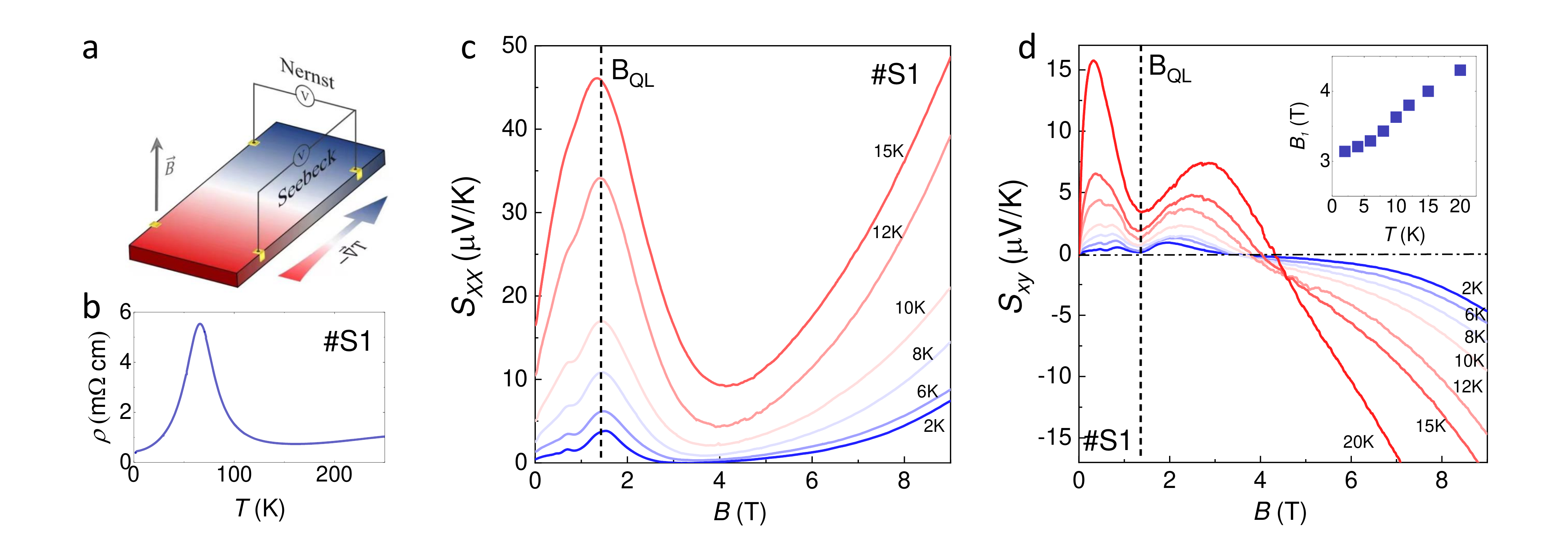}
		\caption{\textbf{Unusual thermoelectric response at $B_{1}$.} (a), Schematic diagram of thermoelectric effect measurements. In a perpendicular magnetic field and a longitudinal thermal gradient, carrier diffusion generates a longitudinal electric field $E_{x}=-S_{xx}|\nabla T|$ (Seebeck effect) and a transverse electric field $E_{y}=S_{xy}|\nabla T|$ (Nernst effect). (b), Temperature dependence of the electrical resistivity ($\rho (T)$) of HfTe$_{5}$ at zero magnetic field. (c), (d), Magnetic field dependence of the thermopower $-S_{xx}$ and Nernst signal $S_{xy}$ at several temperatures below $20~\mathrm{K}$. The vertical dashed line in both diagrams represents the magnetic field at the quantum limit ($B_\mathrm{QL}$). The gap closure in the zeroth Landau band is indicated at $B_{1}$, with $- S_{xx}$ approaching its minimum and $S_{xy}$ undergoing a sign reversal. The inset summarizes $B_{1}$ as a function of temperature.}\label{fig.2}
	\end{figure*}



\noindent  {\fontsize{13}{100}\selectfont \textbf{Gap closing of zeroth Landau bands at $\mathbf{B_1}$}}

\noindent Hereafter, we present our main results in order of increasing magnetic field, starting from zero. Figure \ref{fig.2}b illustrates the temperature-dependent resistivity with a peak at around $T^{*} = 65~\mathrm{K}$. This peak is a typical feature of transition-pentatelluride samples and is attributed to the shift of the Fermi energy from the valence band towards the conduction band as the temperature decreases \cite{Zhouxj17NC}. Thermoelectric experiments were carried out with the thermal gradient along the $a$-axis and the magnetic field along the $b$-axis (see \Fig{fig.2}a). Figure \ref{fig.2}c,d displays the longitudinal (Seebeck) and transverse (Nernst) thermoelectric responses as functions of magnetic field at selected temperatures. At low temperatures, clear quantum oscillations can be identified in both $-S_{xx}$ and $S_{xy}$. The fitted quantum limit for $S1$ is approximately $B_\mathrm{QL} = 1.45~\mathrm{T}$. After the system enters the quantum limit, $-S_{xx}$ exhibits a significant drop, with a magnitude surpassing the quantum oscillations observed at lower magnetic fields. Additionally, this feature can be clearly resolved even at $30~\mathrm{K}$, where the quantum oscillations have completely vanished. Both indicate that the unusual behavior of $-S_{xx}$ and $S_{xy}$ in the quantum limit does not belong to the quantum oscillations as the Fermi level cuts each Landau band.
 
In addition to the drastic drop, the Seebeck signal approaches zero at $B_{1}$, while the Nernst signal exhibits a sign reversal. As discussed preceding \Eq{EqLLBxi}, band inversion and Zeeman effects drive the two zeroth Landau bands to cross each other at the critical field $B_{1}$. The typical transport responses at $B_{1}$ have been investigated in ZrTe$_{5}$ and Pb$_{1-x}$Sn$_{x}$Se \cite{Liangt13NC,Assaf17L,Zhangjl19L}. Qualitatively, our experimentally resolved profile bears a similar appearance to theirs. However, due to the lower carrier density and smaller gap size, the critical field $B_{1}$ of our sample, $3.2~\mathrm{T}$ at $2~\mathrm{K}$, is lower than that in previous studies. As the gap size is enlarged with increasing temperature, warming the sample shifts $B_{1}$ slightly to a higher magnetic field (\Fig{fig.2}c). At low temperatures, two types of electrons contribute to the magnetotransport (see Supplementary Section V). As the temperature increases, $- S_{xx}(B)$ increases from zero to a finite value near $B_{1}$, indicating that the contribution from the trivial pockets becomes non-negligible at higher temperatures. By contrast, the Nernst signal from the trivial pockets—as expected from the Drude model—decays to zero in high magnetic fields. Thus, $S_{xy}(B)$ reflects the intrinsic behavior of the non-trivial electrons in the quantum limit.



\noindent  {\fontsize{13}{100}\selectfont \textbf{Topological Lifshitz transition at $\mathbf{B_2}$}}

\noindent Figure \ref{fig.3}a and b present $-S_{xx}$ and $S_{xy}$ of sample $S2$, measured using a water-cooled magnet up to $33~\mathrm{T}$. The low-field behavior closely matches the results of $S1$. As the magnetic field increases, the thermopower exhibits a continuous rise until reaching its maximum at $B = 11~\mathrm{T}$, after which it decays. This behavior contrasts with theoretical expectations, which suggest that the thermopower of Dirac/Weyl semimetals should grow linearly with the field without saturation \cite{Skinner18SA}. The peak of $-S_{xx}$ in the quantum limit was conjectured to result from a metal-insulator transition, caused by either charge density wave formation or the magnetic freeze-out effect \cite{Tangfd19N,Gourgout22QM}. However, this interpretation has been questioned due to the absence of the thermodynamic evidence expected for a charge density wave transition \cite{Galeski21NC}. Moreover, $\rho_{xx}$ bends above $20 ~\mathrm{T}$ and approaches saturation instead of continuing to diverge as expected for the magnetic freeze-out effect.

Besides the metal-insulator transition, a Lifshitz transition can also produce a peak in $-S_{xx}$ at finite temperature \cite{Varlamov85JETP, Blanter94PR, Varlamov89AP}. As we discussed following \Eq{EqLLBxi}, there is a magnetic-field-induced Lifshitz transition for a weak topological insulator. We take such a mechanism to understand the anomalous behavior of $-S_{xx}$. The excellent agreement between the theoretically calculated $B_2$ (the critical magnetic field for the Lifshitz transition) and our experimentally measured $B_2$ (the peak of $-S_{xx}$) further supports our interpretation. Notably, the temperature dependence of the measured $B_2$ can also be qualitatively explained within this mechanism. 

	\begin{figure*}[tb]\centering
	\includegraphics[width=17.5cm]{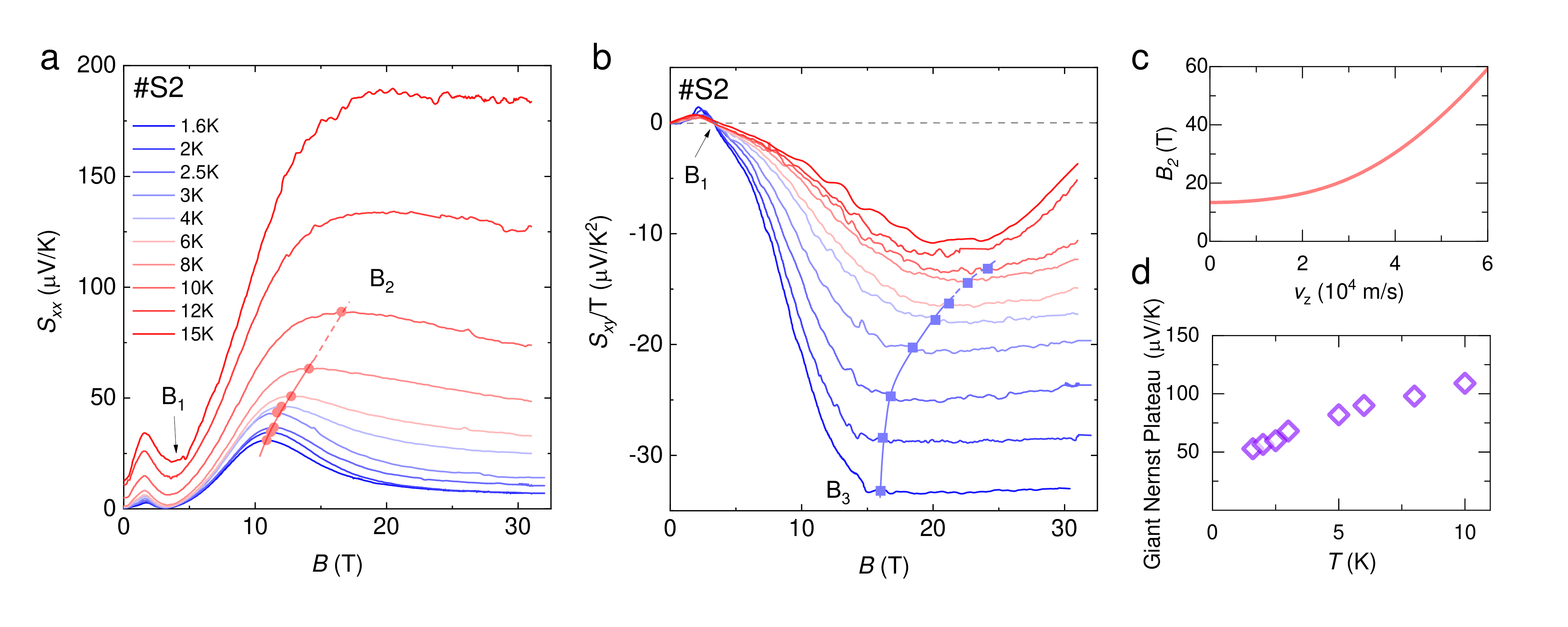}
	\caption{\textbf{Giant Nernst Plateau in ideal 1D Weyl states.} (a), High-field measurements of $-S_{xx}$ up to $33~\mathrm{T}$ at several temperatures. The profiles below $9~\mathrm{T}$ are similar to \Fig{fig.2}c. The critical field $B_2$ for the topological Lifshitz transition is identified from the maximum of $-S_{xx}$ (marked by red circles). $B_2$ shifts to higher magnetic fields with increasing temperature and is not observable above $10~\mathrm{K}$. (b), High-field measurements of the normalized Nernst effect $S_{xy}/T$ at the same temperatures as in (a). A plateau can be observed below $10~\mathrm{K}$. $B_3$ is identified by the field where $S_{xy}/T$ starts to saturate (marked by blue squares). 
    (c), Dependence of $B_{2}$ on $v_z$, calculated using \Eq{EqB2}, with the following parameters: $M_{\perp}=-12.24,\mathrm{eV}\cdot$ \AA$^2$, $M_z=2.7,\mathrm{eV}\cdot$ \AA$^2$, $n=6 \times 10^{17}/\mathrm{cm^3}$, $g=12$, and $\Delta =2.5,\mathrm{meV}$. (d), Summary of the giant Nernst plateau values.}\label{fig.3}
	\end{figure*}
%
 
\begin{figure}[tb]
		\centering
		\includegraphics[width=9cm]{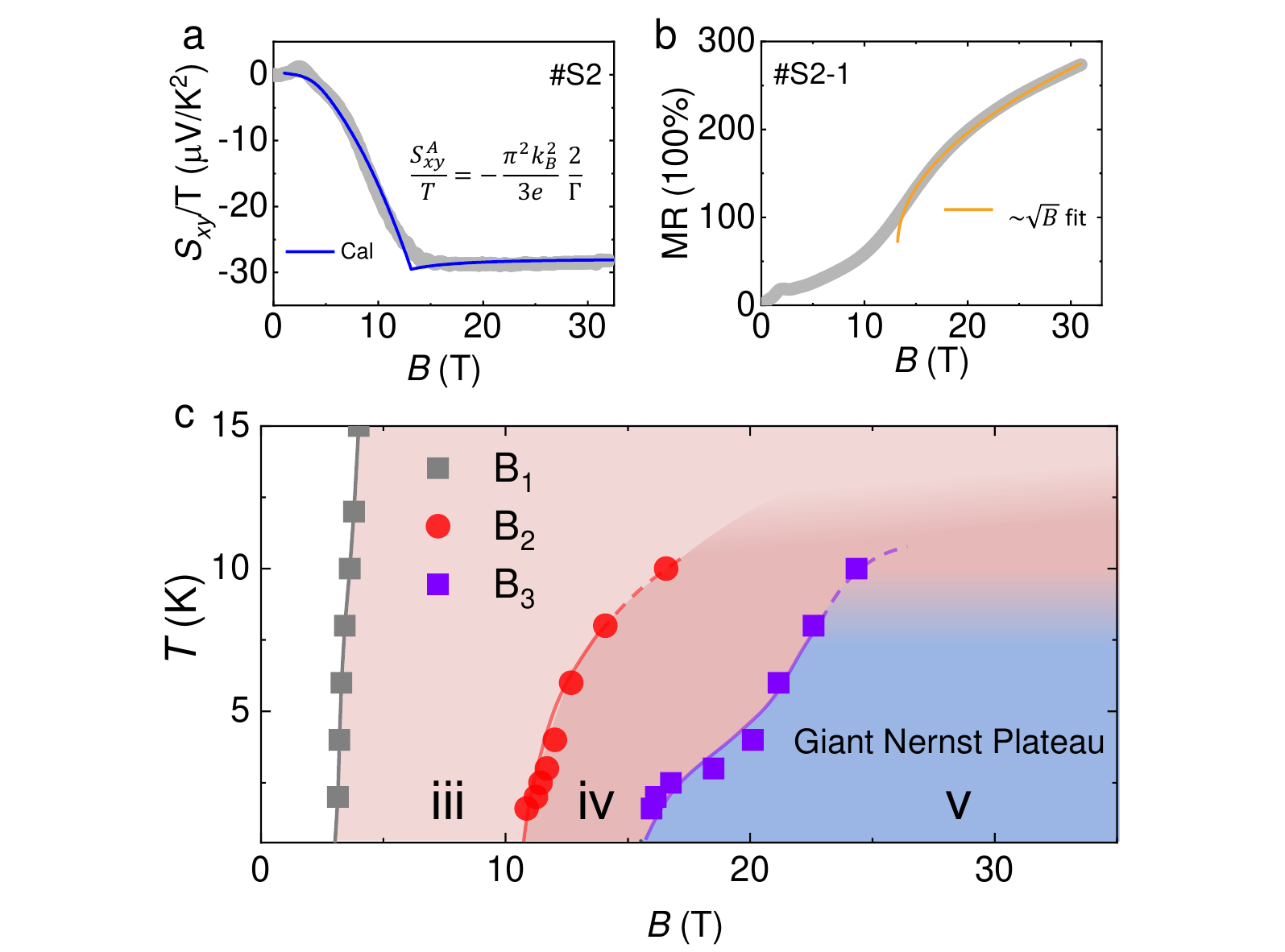}
		\caption{{\textbf{Comparisons of experimental measurements and theoretical calculations.} (a), Experimental (gray line) and calculated $S_{xy}/T$ (blue) at $2~\mathrm{K}$. $S_{xy}/T$ is calculated by fixing $n$ as a constant before the Lifshitz transition ($\approx 13~\mathrm{T}$) and fixing $E_{F}=0$ after the Lifshitz transition (see Supplementary Section VIII for details), with the following parameters: $M_{\perp}=-12.24\,\mathrm{eV}\cdot$ \AA$^2$, $M_z=2.7\,\mathrm{eV}\cdot$ \AA$^2$, $n=8.8 \times 10^{16}/\mathrm{cm^3}, g=12$, and $\Delta =2.5\,\mathrm{meV}$. (b), Magnetoresistance of HfTe$_{5}$ measured at $2~\mathrm{K}$. The yellow line represents a fit with $\sqrt{aB-b}$, where $a$ and $b$ are constants. (c), Temperature–magnetic field phase diagram of HfTe$_{5}$. The values of $B_1$, $B_2$, and $B_3$ are transferred from \Fig{fig.3}. iii, iv, and v denote the corresponding phases in \Fig{fig.1}a.}
	}\label{fig.4}
	\end{figure}

Specifically, $B_{2}$ can be estimated from (see Supplementary Section VIII) 
\begin{eqnarray} 
2M_z\left( \frac{\pi h n}{e} \right)^2-\alpha B_2^3 +\left(\Delta+\frac{\hslash^2 v_z^2}{2M_z}\right)B_2^2= 0, \label{EqB2} \end{eqnarray}
which indicates that $B_{2}\sim n^{2/3}$ and $B_{2}\sim v_z^2$. The carrier density of our sample in zero field is approximately $n_{0} = 6.0 \times 10^{17}/\mathrm{cm}^{3}$. By substituting this value into \Eq{EqB2} with $v_z=0$, the calculated value is approximately $13~\mathrm{T}$, which is consistent with our experimental results. 
The previous ARPES experiment \cite{Zhouxj17NC} indicates that as the temperature increases, the zero-field Fermi level moves closer to the Dirac point, making the contribution of the linear band more significant. Thus, $v_z$ increases as the temperature rises. Taking $v_z(15~\mathrm{K})=6\times 10^{4}\mathrm{m/s}$ with the same carrier density results in $B_2\approx 60~\mathrm{T}$ (\Fig{fig.4}c), which is beyond the accessible range of our study. In the case without such a topological Lifshitz transition, $-S_{xx}$ exhibits saturating behavior, as expected in a trivial semimetal. The excellent agreement between the theoretically calculated $B_2$ (the critical magnetic field for the Lifshitz transition) and our experimentally measured $B_2$ (the peak of $-S_{xx}$) further supports our interpretation.

\noindent  {\fontsize{13}{100}\selectfont \textbf{Giant Nernst plateau above $\mathbf{B_3}$}}

\noindent We now present the high-field Nernst plateau shown in \Fig{fig.3}b. Notably, at low temperatures, after an initial decrease, $S_{xy}$ develops an extended plateau beyond a saturation field. The saturating behavior of $S_{xy}$ indicates the presence of an anomalous, i.e., a magnetic-field-independent component in the Nernst signal. Such an anomalous term can reach an ultra-high value of $50~\mu\mathrm{V}/\mathrm{K}$ at $2~\mathrm{K}$. As the temperature increases, the onset of the plateau moves to a higher magnetic field and its height increases (see \Fig{fig.3}d), reaching $110~\mu\mathrm{V}/\mathrm{K}$ at the maximum temperature of $10~\mathrm{K}$, where the plateau remains clearly visible. Recently, the presence of a non-zero Berry curvature from the spin-split massive Dirac bands has been proposed to explain the high-field ANE in non-magnetic topological materials \cite{Liangt17L,Caglieris18B,Liuyz21B,Sunzl20QM,Gourgout22QM}. However, it may not apply in this case because Berry curvature is absent in 1D Landau bands. As discussed in \Fig{fig.1}, carrier density above $B_{2}$ is dramatically reduced due to the compensation between electron carriers and hole carriers from $E_{0-}$. As a result, the higher magnetic fields would immediately push the Fermi energy toward the two 1D Weyl points. Since the high-field anomalous Nernst response exists exclusively above $B_{2}$, it implies a strong correlation with the ideal 1D Weyl phase.

We theoretically demonstrate that the observed plateau in $S_{xy}$ is caused by the ideal 1D Weyl phase shown in \Fig{fig.1}e. According to the Mott relation, $S_{x y}=L_0e (\rho_{xx} \partial \sigma_{xy}/\partial E_F-\rho_{yx}\partial \sigma_{xx}/\partial E_F)$, where $L_0=\pi^2 k_B^2 T/3 e^2$ denotes the Lorentz number. The conductivity meets the requirements $\sigma_{xy}(-E_F)=-\sigma_{xy}(E_F)$ and $\sigma_{xx}(-E_F)=\sigma_{xx}(E_F)$. Thus, in the ideal 1D Weyl phase where $\left. \partial \sigma_{xx} /\partial E_F \right|_{E_F \rightarrow 0}=0$, the Nernst cofficient is determined by $S_{xy}=L_0e \sigma_{xx}^{-1} \partial \sigma_{xy}/\partial E_F$, where $\sigma_{xy}$ contains only the Drude term and the anomalous Hall is not included. Based on the Kubo formula, we obtain $\partial\sigma_{xy}/\partial E_F = -2/\left(\pi R_{\mathrm{K}}v_{w}\right)$ and $\sigma_{xx} = \Gamma/\left(\pi R_{\mathrm{K}}v_{w}\right)$ (see Supplementary Section IX), where $R_{\mathrm{K}}=h/e^2$ denotes the von Klitzing constant, $v_w=2\sqrt{M_z\left(\alpha B-\Delta\right)}$ denotes the Fermi velocity of the ideal 1D Weyl phase, and $\Gamma$ is the energy broadening of the Landau bands. As a result, the Nernst coefficient is given by
		\begin{equation}
		\frac{S_{x y}}{T} =-\frac{\pi^2 k_B^2 }{3e} \frac{2}{\Gamma}. \label{EqSxy}
	\end{equation}
Apart from the basic constants $\pi^2 k_B^2/3e$, the Nernst coefficient is entirely determined by the energy broadening parameter $\Gamma$, which is field-independent in the ultra-quantum limit \cite{Gornik85L,Smith85B,Ashoori92SSC}.
By taking $\Gamma=2\,\mathrm{meV}\sim \Delta$, the plateau value is obtained as $S_{xy}/T=-24.4\,\mu\mathrm{V}/\mathrm{K^2}$, which is comparable to the experimentally measured values. For a higher temperature, a larger $\Gamma$ is expected 
\cite{Weihp85B}
 and thus a lower $S_{xy}/T$ is achieved. Our complete theory not only explains the origin of the plateau but also predicts the behavior of $S_{xy}(B)$ in the quantum limit, which is in excellent agreement with experimental results, as shown in \Fig{fig.4}a. Moreover, $\rho_{xx}=\sigma_{xx}^{-1} = \left(\pi R_{\mathrm{K}}v_{w}\right)/\Gamma$ in the ideal 1D Weyl phase suggests that $\rho_{xx}$ increases with the magnetic field as $\sqrt{a B - b}$. The measured magnetoresistance ($(\rho_{xx}(B)-\rho_{xx}(0))/\rho_{xx}(0)$) shown in \Fig{fig.4}b can be well fitted by this expression when $B > 15~\mathrm{T}$ (ideal 1D Weyl phase), which further confirms the validity of our explanations.

\noindent  {\fontsize{13}{100}\selectfont \textbf{Discussions}}

\noindent We summarize our findings with a phase diagram in the $B$–$T$ plane, shown in \Fig{fig.4}c. Phenomenally, the giant Nernst plateau observed here is quite similar to the widely reported ANE \cite{Nagaosa10RMP}. However, they differ fundamentally in both experimental measurements and theoretical interpretations. Firstly, the giant Nernst plateau occurs at an ultra-quantum limit, while ANE is a low-field effect. To our knowledge, no experiments have observed plateaus in such strong magnetic fields.
Secondly, the giant Nernst plateau represents the new mechanism of thermoelectric conversion, where the plateau is quite different from the semi-classical theories for ANE \cite{Nagaosa10RMP,Behnia16RPP}.
Notably, such a plateau is unrelated to the anomalous Hall conductivity, since our theoretical calculations show that the Hall conductivity contains only the Drude term after the formation of Landau bands (Supplementary Section VII). Finally, the giant Nernst plateau ($S_{xy}/T$) is exceptionally large, surpassing any measured ANEs to date \cite{Ikhlas17NP,Sakai18NP,Watzman18B,Yanghy20PRM,Asaba21SA,Chents22SA,Pany22NM}, and \Eq{EqSxy} indicates that increasing mobility can further enhance its magnitude. 
Moreover, reducing the carrier density will significantly lower the magnetic field required to form the ideal 1D Weyl phase. 
In this sense, the weak topological insulators with extremely low quantum limit have potential applications in the development of novel thermoelectric devices.





\noindent  {\fontsize{13}{100}\selectfont \textbf{Acknowledgements}}

\noindent 
We acknowledge very helpful discussions with X. Yuan, C. L. Zhang, C. Zhang and Z. J. Xiang. This work was financially supported by the National Key R$\&$D Program of the MOST of China (Grant No. 2022YFA1602602, 2022YFA1602603, and 2022YFA1403700), the National Natural Science Foundation of China (Grants No. 12474053, 12122411, 12104459, 12304074, 12234017, 12374041, 11925402, 12004157, and 12350402), the Basic Research Program of the Chinese Academy of Sciences Based on Major Scientific Infrastructures (Grants No.JZHKYPT-2021-08), Excellent Program of Hefei Science Center CAS (Grant No. 2021HSC-CIP016), The { CASHIPS Director' Fund} (Grant No. YZJJ2022QN36), Guangdong Basic and Applied Basic Research Foundation (2023B0303000011), Guangdong province (2020KCXTD001), the Science, Technology and Innovation Commission of Shenzhen Municipality (ZDSYS20190902092905285), and Center for Computational Science and Engineering of SUSTech. 

\noindent  {\fontsize{13}{100}\selectfont \textbf{Author contributions}}

\noindent 
Y.L. and Y.C. synthetized the HfTe$_5$ crystals.
Y.Z., Q.L. and Y.Q. performed the thermoelectric transport measurements. Y.Z., Q.L.,P.Z., J.Z. analysed the data. P.Z. performed the theoretical analysis with assistance from H.L. and Q.N.. J.Z., P.Z. and Q.L.wrote the paper with help from all other co-authors. J.Z., M.T. and H.L. supervised the project. All authors discussed the results and commented on the manuscript.

\noindent  {\fontsize{13}{100}\selectfont \textbf{Competing interests}}

\noindent  The authors declare no competing interests.

	
\bibliographystyle{apsrev4-1-etal-title_6authors}
\bibliography{HfTe5_20241012.bib}

\end{document}